\documentclass[%
aapm,
mph,%
amsmath,amssymb,
 reprint,%
]{revtex4-1}

\usepackage{graphicx}
\usepackage{dcolumn}
\usepackage{bm}
\usepackage{xcolor}
\usepackage{xfrac}
\usepackage{lmodern}
\usepackage{type1cm} 
\usepackage{enumitem}
\usepackage{graphicx}
\usepackage{dcolumn}
\usepackage{bm}

\usepackage[mathlines]{lineno}
\modulolinenumbers[5]

\begin{document}


\def\HH{{\cal H}}

\title{Observation of fractional  evolution in nonlinear optics}%



\author{Van Thuy Hoang}
\email{vanthuy.hoang@sydney.edu.au}
\author{Justin Widjaja}%
\altaffiliation[Current address ]{Department of Applied Physics and Materials Science, California Institute of Technology, 1200 E. California Boulevard, Pasadena, CA 91125, USA}
\author{Y. Long Qiang}%
\affiliation{ 
Institute of Photonics and Optical Science (IPOS), School of Physics A28, The University of Sydney, Sydney, 2006, New South Wales, Australia 
}%
\affiliation{ARC Centre of Excellence for Optical Microcombs for BreakthroughScience (COMBS)} 

\author{Maxwell Liu}%
\author{Tristram J. Alexander}%
\affiliation{ 
Institute of Photonics and Optical Science (IPOS), School of Physics A28, The University of Sydney, Sydney, 2006, New South Wales, Australia 
}%

\author{Antoine F. J. Runge}%
\author{C. Martijn de Sterke}%
\affiliation{ 
Institute of Photonics and Optical Science (IPOS), School of Physics A28, The University of Sydney, Sydney, 2006, New South Wales, Australia 
}%
\affiliation{ARC Centre of Excellence for Optical Microcombs for BreakthroughScience (COMBS)}

\begin{abstract}
The idea of fractional derivatives has a long history that dates back centuries. Apart from their intriguing mathematical properties, fractional derivatives have been studied widely in physics, for example in quantum mechanics and generally in systems with nonlocal temporal or spatial interactions. However, systematic experiments have been rare due to challenges associated with the physical implementation. Here we report the observation and full characterization of a family of temporal optical solitons that are governed by a nonlinear wave equation with a fractional Laplacian. This equation has solutions with unique properties such as non-exponential tails and a very small time-bandwidth product.
\end{abstract}

\maketitle

\section{Introduction}\label{sec:intro}

Derivatives, as a fundamental concept of calculus for more than three centuries, are usually understood to be of integer order, however a more general class of fractional order derivatives can be defined \cite{Herrmann_Fractional}. The idea of a fractional derivative was raised as early as 1695 in a correspondence between Leibniz and L'Hospital \cite{
Ross_1977}, with fractional calculus being derived multiple times in attempts to understand physical observations, from the early work of Abel~\cite{Podlubny2017} to the formalism of Caputo \cite{Caputo1967}.  

Fractional derivatives are generally computed through integrals that depend on the values of a function over an entire interval ~\cite{Vasquez_2011}, so the equations they enter are said to be {\sl nonlocal}. Fractional derivatives are thus well suited to describe nonlocal physical phenomena, either in time (memory) or in space (long-range interaction). They have been investigated in physical systems such as ferromagnetism \cite{Kovalev_1986}, nonlocal elasticity \cite{Carpinteri_2011, Carpinteri_2014},  
phase separation \cite{Akagi_2016}, ultrasound, \cite{Chen_2004}
porous media \cite{Depablo_2011}, anomalous diffusion \cite{Yamamoto_2012}, and non-diffusive transport in plasmas \cite{Castillo-Negrete_2006}. However, it is a recent generalization of quantum mechanics to the fractional case~\cite{Laskin_2000} that has led to an increased interest in fractional derivative models. 

In fractional quantum mechanics, a fractional Laplacian enters as a (fractional) kinetic energy operator~\cite{Laskin_2000, Laskin2_2000, Laskin_2002} when replacing the statistics of Brownian motion in the calculation of path integrals by that of L\'evy flights \cite{Laskin_2000,Malomed_2024}.  This leads to the fractional Laplacian operator $(-\nabla^2)^{\alpha/2}$, with the L\'evy index $0<\alpha<2$ ~\cite{Herrmann_Fractional, Mandelbrot_Fractal, Malomed_2024}, which reverts to the usual Laplacian when $\alpha=2$.  L\'evy flights can be found throughout science, including, for example, in earthquake statistics \cite{Sotolongo_2000} and in the search strategies of predators \cite{Humphries_2010}. Even though fractional Laplacians have been defined in as many as $10$ different ways, it has been shown that all of these are equivalent  \cite{Kwasnicki_2017}.

The Laplacian operator is fundamental to wave models, including the widely used nonlinear Schr\"{o}dinger (NLS) equation.  The generalization of the Laplacian to the fractional case, and therefore the emergence of a fractional nonlinear Schr\"{o}dinger equation, has prompted significant interest in fractional nonlinear wave theory.  Rogue waves~\cite{Rizvi2021}, optical vortex solitons~\cite{Li2020}, spin-orbit coupled solitons in Bose-Einstein condensates~\cite{Sakaguchi2022}, and dissipative solitons~\cite{Qiu2020} have all been found in fractional nonlinear wave equations. In applied mathematics, the nature of the ground state \cite{Frank_2013,Seok_2021}, the well-posedness of the equations~\cite{Guo_2010,Sy_2022}, and the ``blow-up'' of the solutions~\cite{Boulenger_2016,Klein_2014}, depending on the parameter $\alpha$, have all been studied.  However, despite this considerable interest from multiple, large communities, the absence of associated experiments is striking.

\begin{figure*}[ht]
\centering
\includegraphics[width=13cm]{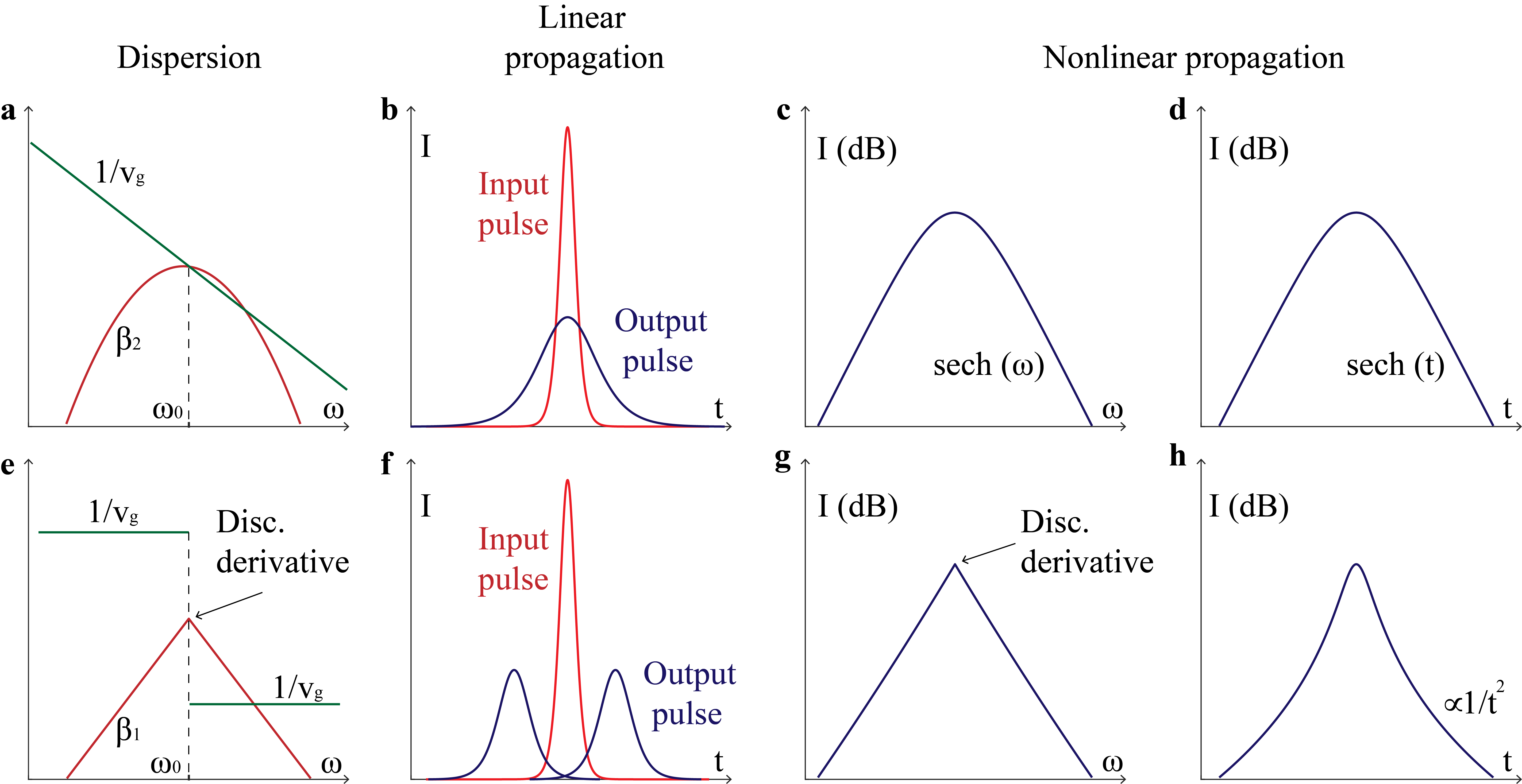}
\caption{\textbf{Concept of Hilbert-NLS soliton.} (a)-(d) Conventional solitons can form in the presence of quadratic dispersion. (a) The dispersion (red) and (inverse) group velocity (green) vary smoothly with frequency; (b) At low intensities,  dispersion stretches pulses in time. Solitons can form at high intensities with smooth (c) spectral and (d) temporal profiles that decay exponentially. Hilbert-NLS solitons (e) arise in the presence of dispersion relation Eq.~\eqref{eq:disp_rel} which has a discontinuous derivative (red) and associated discontinuous (inverse) group velocity (green) as a function of frequency; (f) At low intensities the dispersion causes input pulses to split in two. At high intensities solitons form which have (g) a spectrum with discontinuous derivative, and (h) non-exponential decay in time.}
\label{fig:concept}
\end{figure*}
Optics is one of the most promising fields for the experimental realisation of physical systems with fractional Laplacians, because of its mature technology and the direct link between the model equations and the underlying physics. Indeed, the {\sl linear} fractional propagation of light beams has already been studied theoretically \cite{Longhi_2015, Zhang_2015, Zhang_2016}, and the first experimental demonstration was recently reported \cite{Liu_2023}. However, these experiments considered free propagation, which, in the context of quantum mechanics, corresponds to the simple case of evolution without an external potential. In fact, any experimental investigations of a fractional Laplacian with a potential, either linear or nonlinear, have yet to be reported.

Here we provide such a demonstration: we report the generation and characterization of temporal solitons governed by a fractional Laplacian. The experiments are carried out in a fibre laser in which the nonlocal temporal effects can be programmed through the dispersion: in a co-moving frame the dependence of the propagation constant of the light $\beta$ on the frequency $\omega$ \cite{Runge_2020}. We consider a {\sl dispersion relation} of the form 
\begin{equation}
\beta(\omega) = -|\beta_1||(\omega-\omega_0)|,    
\label{eq:disp_rel}
\end{equation}
where $\omega_0$ is the central frequency. The inverse group velocity $v_g^{-1}$ (in the co-moving frame) is related to the wavenumber by $v_g^{-1}=d\beta/d\omega$. Thus for $\omega<\omega_0$, $v_g=1/|\beta_1|$, whereas for $\omega>\omega_0$, $v_g=-1/|\beta_1|$. This dispersion relation and associated inverse group velocity are illustrated in Fig.~\ref{fig:concept}(e), and should be compared to the parabolic dispersion relation for conventional NLS solitons in Fig.~\ref{fig:concept}(a). At low intensities, when nonlinear effects are negligible, dispersion relation Eq.~\eqref{eq:disp_rel} causes the pulse to split in two parts, with frequencies $\omega>\omega_0$ speeding up and the frequencies $\omega<\omega_0$ slowing down (Fig.~\ref{fig:concept}(f)), as previously demonstrated by Liu \textit{et al.} \cite{Liu_2023}. In contrast, in the conventional case the continuous variations of the group velocity lead to pulse broadening (Fig.~\ref{fig:concept}(b)). 

Combining the dispersion with the nonlinear properties of the optical fibre can lead to solitons. In the conventional case, described by the NLS equation, these solitons have a hyperbolic secant shape, both in frequency (Fig.~\ref{fig:concept}(c)) and in time (Fig.~\ref{fig:concept}(d)). They therefore are smooth and decay exponentially in both domains. In contrast, dispersion relation following Eq.~\eqref{eq:disp_rel} leads to solitons with a discontinuous derivative in frequency  (Fig.~\ref{fig:concept}(g)), and which decay algebraically in time (Fig.~\ref{fig:concept}(h)). These solitons satisfy the Hilbert-nonlinear Schr\"odinger (Hilbert-NLS) equation Eq.~\eqref{eq:HNLSE} \cite{Gaididei_1997} to be discussed in Section~\ref{sec:concept}. This equation is a special case of the fractional NLS equation \cite{Kirkpatrick_2016}, and has fractional Laplacian with $\alpha=1$. 

Using spectral, temporal, and phase-resolved measurements, we demonstrate the unique properties of the Hilbert-NLS solitons and compare these with the properties of the conventional solitons of the NLS equation for which $\alpha=2$. In particular, we explore (i) the intrinsic properties of the soliton shape (Section~\ref{sec:shape}); 
(ii) the pulse-width dependence of the soliton energy at fixed $\beta_1$ (Section~\ref{sec:fixed_beta_1}); 
and
(iii) the soliton energy as $\beta_1$ is varied (Section~\ref{sec:varying_beta_1}). 
Although we focus on the case with $\alpha=1$, the flexibility of our experimental apparatus allows for the generation of solitons at any value of the L\'evy index $\alpha$, and we demonstrate the cases $\alpha = \sfrac{5}{4}$ and $\alpha = \sfrac{\pi}{2}$  (Section~\ref{sec:other_alpha}). 

Taken together, these results demonstrate that fractional nonlinear phenomena can be demonstrated experimentally--they pave the way for future experiments with new analogies in quantum mechanics and many other areas of physics and applied mathematics. 

\section{Hilbert-nonlinear Schr\"odinger equation}
\label{sec:concept}

The propagation of light in a nonlinear medium with dispersion relation Eq.~\eqref{eq:disp_rel} is governed by the Hilbert-NLS equation \cite{Gaididei_1997} (see Methods)
\begin{equation}
i\frac{\partial\varphi}{\partial z}-|\beta_1|~\HH\!\left(\frac{\partial\varphi}{\partial t}\right)+\gamma |\varphi|^2\varphi=0,
\label{eq:HNLSE}
\end{equation}
where $\varphi(z,t)$ is the electric field envelope, $z$ is the propagation distance, $t$ is time, and $\gamma$ is the nonlinear coefficient \cite{Kivshar}. The operator ${\cal H}$ denotes the Hilbert transform  \cite{King}, an integral, (i.e. nonlocal) operator. The Hilbert transform, in essence, reverses the sign of the amplitudes of the negative frequency components of its argument, and in this way it thus expresses the absolute value that enters Eq.~\eqref{eq:disp_rel}. The operator in the second term in Eq.~\eqref{eq:HNLSE} corresponds to the fractional Laplacian with $\alpha=1$ according to Riesz \cite{Kwasnicki_2017}. Equation~\eqref{eq:HNLSE} also arises in the study of ferromagnetism \cite{Kovalev_1986} and in the quasi-continuum limit of investigations of nonlinear, one-dimensional systems with long-range dispersive interactions \cite{Gaididei_1997}.

In our experiments, we generate optical pulses that have an intensity that remains unchanged upon propagation \cite{Runge_2020,Runge_2021}. Such {\sl stationary} solutions $u(t)$ of {Eq.~\eqref{eq:HNLSE} are defined through $\varphi(z,t)=u(t) e^{i\mu z}$---where $\mu$ is a nonlinear contribution to the propagation constant---and satisfy 
\begin{equation}
-\mu u-|\beta_1|~\HH\!\left(\frac{\partial u}{\partial t}\right)+\gamma u^3=0.
\label{eq:stationary}
\end{equation}
Although analytic solutions to Eq.~\eqref{eq:stationary} are not known, from its general properties, combined with numerical solutions \cite{Yang_NLW}, we can find that these solitons have unique features that differ substantially from those of the conventional hyperbolic secant solitons \cite{Kivshar}.

To highlight the difference between Eq.~\eqref{eq:HNLSE} and the conventional NLS equation we consider their scaling relations. For example, it is straightforward to see that if $\varphi(t,z)=u(t)e^{i\mu z}$ is a solution of Eq.~\eqref{eq:HNLSE}, then so is $\tilde\varphi(t,z)=\eta~u(\eta^2 t)~ e^{i\eta^2\mu z}$, where $\eta$ is real and positive. This relies on the property of Hilbert transforms that if $g(t) = {\cal H}(f(t))$, then ${\cal H}(f(at)) = g(at)$ for any real, positive $a$ \cite{King}. The scaling relation between $\varphi$ and $\tilde\varphi$ means that the pulse energy $\int|\varphi|^2dt$ is independent of $\mu$; the pulse energy thus does not depend on the pulse length at constant $\beta_1$ and $\gamma$ and it therefore is, in effect, quantized. From this, and other such relations (see Methods) it can be shown that 
\begin{equation}
    E=K\frac{|\beta_1|}{\gamma},
\label{eq:scaling_K}
\end{equation}
where $K$ is a constant. From numerical calculations, it is found to take the value $K = 2.47$. This is a firm prediction which we confirm experimentally in Section~\ref{sec:fixed_beta_1}.
It contrasts strongly with conventional solitons for which the pulse energy $E\propto1/\Delta t$, where $\Delta t$ is the pulse length. 

Before turning to our experimental results, we note that according to the Vakhitov-Kolokolov criterion \cite{Malomed_2024, Vakhitov_1973,Boulenger_2016}, Eq.~\eqref{eq:stationary} is marginally stable, i.e. it is on the boundary between being stable and being unstable. This means that, depending on the initial condition, nominally stationary solutions either disperse or diverge (``blow up''). This behavior can be arrested by applying a small additional perturbation \cite{Gaididei_1999, 
Stephanovich_2022}, as discussed in Section~\ref{sec:Exp}.

\section{Experimental results}\label{sec:Exp}

Our experiments are performed using a passively mode-locked fibre laser that incorporates an intracavity pulse shaper \cite{Runge_2020, Mao_2021}. The pulse shaper applies a phase that cancels the second- and third-order dispersion introduced by the optical fibre, so that the effects of these orders are negligible, and it also generates a dominant phase as $\phi(\omega) =-|\beta_1||\omega-\omega_0|L$, where $L$ is the cavity length, consistent with Eq.~\eqref{eq:disp_rel} (see Supplementary Information 
). The output pulses are characterised through a set of temporal and phase-resolved measurements using a frequency-resolved electrical gating (FREG) apparatus (see Methods) \cite{Dorrer_2002}.

To address the marginal stability of these solitons (Sec.~\ref{sec:concept})
we add a small amount of negative quartic dispersion at frequencies far from the central frequency (see Methods) where the spectral soliton amplitude is low (orange curve in Fig.~\ref{fig:spectrum}(a)). However, the effect of this quartic dispersion on the generated soliton remains small in our experiments as discussed below.

\subsection{Soliton shape at $\alpha=1$}
\label{sec:shape}

For the results discussed here, we first apply the phase profile shown in Fig.~\ref{fig:spectrum}(a) (solid orange) for which $|\beta_1| = 40~\rm{ps~km^{-1}}$. The corresponding output spectrum of the laser operating in the Hilbert-NLS regime is shown in Fig.~\ref{fig:spectrum} (solid blue). It exhibits a triangular-like shape on a logarithmic scale, and is in very good agreement with the predicted shape (red dashed), calculated from solving Eq.~\eqref{eq:stationary} numerically \cite{Yang_NLW}. Note the discontinuous derivative at the central frequency $\omega_0$, a consequence of the discontinuous derivative at the origin in Eq.~\eqref{eq:disp_rel}. The sharp features on both sides of the spectrum are Kelly sidebands and can be ignored ~\cite{Kelly_1992}.

\begin{figure}[h]
\centering
\includegraphics[width=8cm]{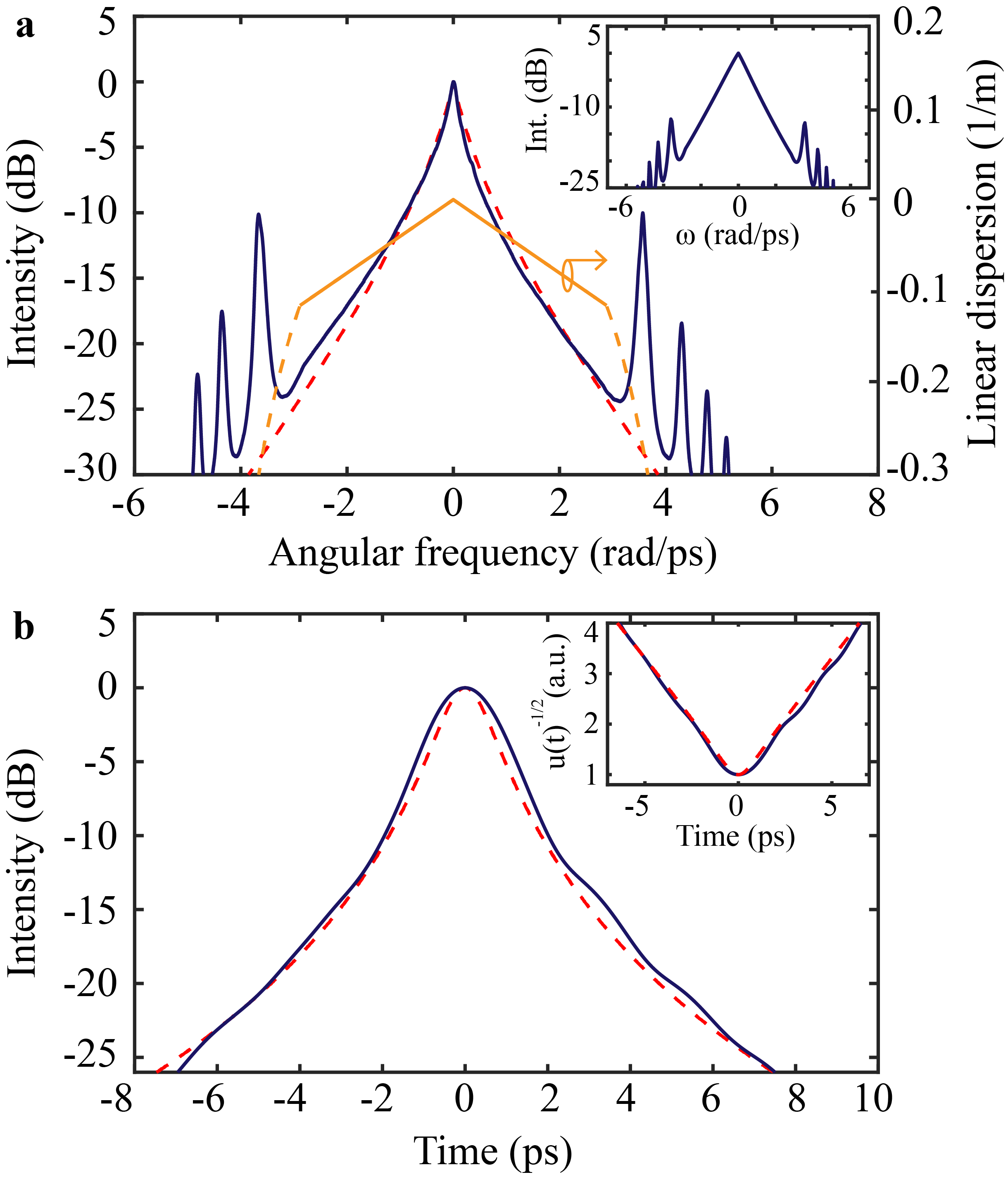}
\vskip -1mm
\caption{\textbf{Spectro-temporal characterization of the Hilbert-NLS soliton regime.} (a) Measured (solid blue) and numerical (red dashed) spectra for $|\beta_1| = 40~\rm{ps~km^{-1}}$. The orange curve shows the net dispersion of the cavity; the solid curve corresponds to Eq.~\eqref{eq:disp_rel}, whereas the dashed curve corresponds to the quartic dispersion included to provide stability. The inset shows the simulated laser output spectrum; (b) Retrieved (solid blue) and numerically calculated (red dashed) temporal intensity profiles. The inset shows $(u(t))^{-1/2}$ versus time.}
\label{fig:spectrum}
\end{figure}

We also modeled the laser dynamics using an iterative cavity map (see Supplementary Information 
) \cite{Runge_2020}. The simulated output spectrum of the laser, for the same parameters, is shown in the inset of Fig.~\ref{fig:spectrum}(a). It is in good agreement with the experiments and is consistent with the change of pulse-shaping mechanism.

The discontinuous derivative in the spectrum causes $u(t)$ to be proportional to $\sfrac{1}{t^2}$ as $|t|\rightarrow\infty$ and is a manifestation of the nonlocal nature of the fractional Laplacian; the solutions thus decay algebraically unlike the exponential decay of conventional solitons \cite{Kivshar}. This is confirmed in Fig.~\ref{fig:spectrum}(b), where the retrieved temporal intensity (solid blue) decays sublinearly on a logarthmic scale. This indicates slower-than-exponential decay, again in very good agreement with the predicted shape (red dashed). The small oscillations in the retrieved temporal shape arise from the quartic dispersion that is introduced to suppress the instabilities (see Supplementary Information 
). The inset of Fig.~\ref{fig:spectrum}(b) shows the same data but with $u(t)^{-\sfrac{1}{2}}$ on the vertical axis. Presented this way, the data is a straight line far from the origin, consistent with the expected algebraic decay.

The time-bandwidth product $\Delta f\hspace{0.5mm}\Delta t$---the product of the full-width at half maximum (FWHM) of the spectral and temporal intensities---is characteristic to the particular pulse shape. For conventional hyperbolic secant solitons, for example, it takes the value $0.315$ \cite{Kivshar}. For the soliton solutions to Eq.~\eqref{eq:stationary}, it takes the extremely small value $\Delta f\hspace{0.5mm}\Delta t=0.063$ based on numerical calculations. By measuring the FWHM of the pulses in Fig.~\ref{fig:spectrum}(a) and (b), we find $\Delta f=0.046~{\rm THz}$ and $\Delta t=1.81~{\rm ps}$, corresponding to a time-bandwidth product of $0.083$. The fact that this is almost four times smaller than for hyperbolic secant pulses highlights the marked difference between the pulses we generate and conventional solitons. The small discrepancy of the measured time-bandwidth product with the theoretical value is likely to arise from residual chirp in our dispersion-managed laser configuration \cite{Runge_2020}.

\subsection{Soliton energy at $\alpha=1$ with fixed $\beta_1$}\label{sec:fixed_beta_1}

As discussed in 
Sec.~\ref{sec:concept}, 
the energy $E$ of the soliton solutions of Eq.~\eqref{eq:HNLSE} is given by Eq.~\eqref{eq:scaling_K}, and is independent of the pulse duration $\Delta t$. We verify this prediction by measuring the output pulse duration and energy for different pump powers. We deduct the portion of the energy in the spectral sidebands by integrating the corresponding measured optical spectrum. Results of these measurements are shown in Fig.~\ref{fig:energy}(a) (blue circles), and are in good agreement with the predicted soliton energy calculated from Eq.~\eqref{eq:scaling_K} for the same parameters (solid red) once we account for the output coupling and insertion loss of the spectral shaper \cite{Runge_2020}. The measured output energy varies by only approximately $17\%$ as the pulse width varies by a factor $3$. On the other hand, the expected energy of conventional solitons for the parameters of our laser cavity, is shown by the solid green curve and follows a $\sfrac{1}{\Delta t}$-dependence (see Methods) \cite{Runge_2020,Kivshar}. For these conventional pulses, therefore, the pulse energy changes by a factor $3$ as the pulse duration varies by the same factor. The observation that the soliton energy is approximately constant as the pulse width varies by a factor $3$ indicates once more that the clear difference the solitons we generate and conventional NLS solitons. We note that the contrast with some other solitons is even more striking; for example, for pure quartic solitons the energy is proportional to $(\Delta t)^{-3}$~\cite{Runge_2020}. In experiments such as in Fig.~\ref{fig:energy}(a), their energy thus varies by a factor $27$ as the pulse length varies by a factor $3$ in contrast with the approximate constant energy of Hilbert-NLS solitons.  

For larger pump powers, the laser cavity can sustain several solitons propagating at the same time. As Eq.~\eqref{eq:scaling_K} predicts that the energy of each pulse is equal, the total soliton energy in the cavity jumps as the pump power is changed and the number of solitons in the cavity changes. We measure the output pulse train using a slow photodetector and oscilloscope for up to $n_p = 5$ solitons in the laser cavity. The energy of each pulse is then calculated by integrating the recorded temporal trace. In all cases, we find that, (i) the energy of each soliton is consistent with Fig.~\ref{fig:energy}(a) and is approximately constant over the pump power range for which a given number of pulses exist in the cavity (see Supplementary Information 
); and (ii) the energy of the total number $n_p$ of solitons is $n_p$ times the energy of a single soliton. Experimental confirmation is provided in Fig.~\ref{fig:energy}(b). It shows that the output energy of the laser jumps by approximately $52~{\rm pJ}$ as an additional soliton enters the cavity, consistent with Eq.~\eqref{eq:scaling_K}.  

We note that the quantization of the pulse energy is not novel in itself. Grudinin \textit{et al.} \cite{Grudinin_1992} for example, reported a laser with this property. However, in their case, the fixed energy arises from the properties of the gain medium and the mode-locker. In contrast, in our work, it is entirely due to phase properties of the cavity. 

\begin{figure}[ht!]
\centering
\includegraphics[width=7.5cm]{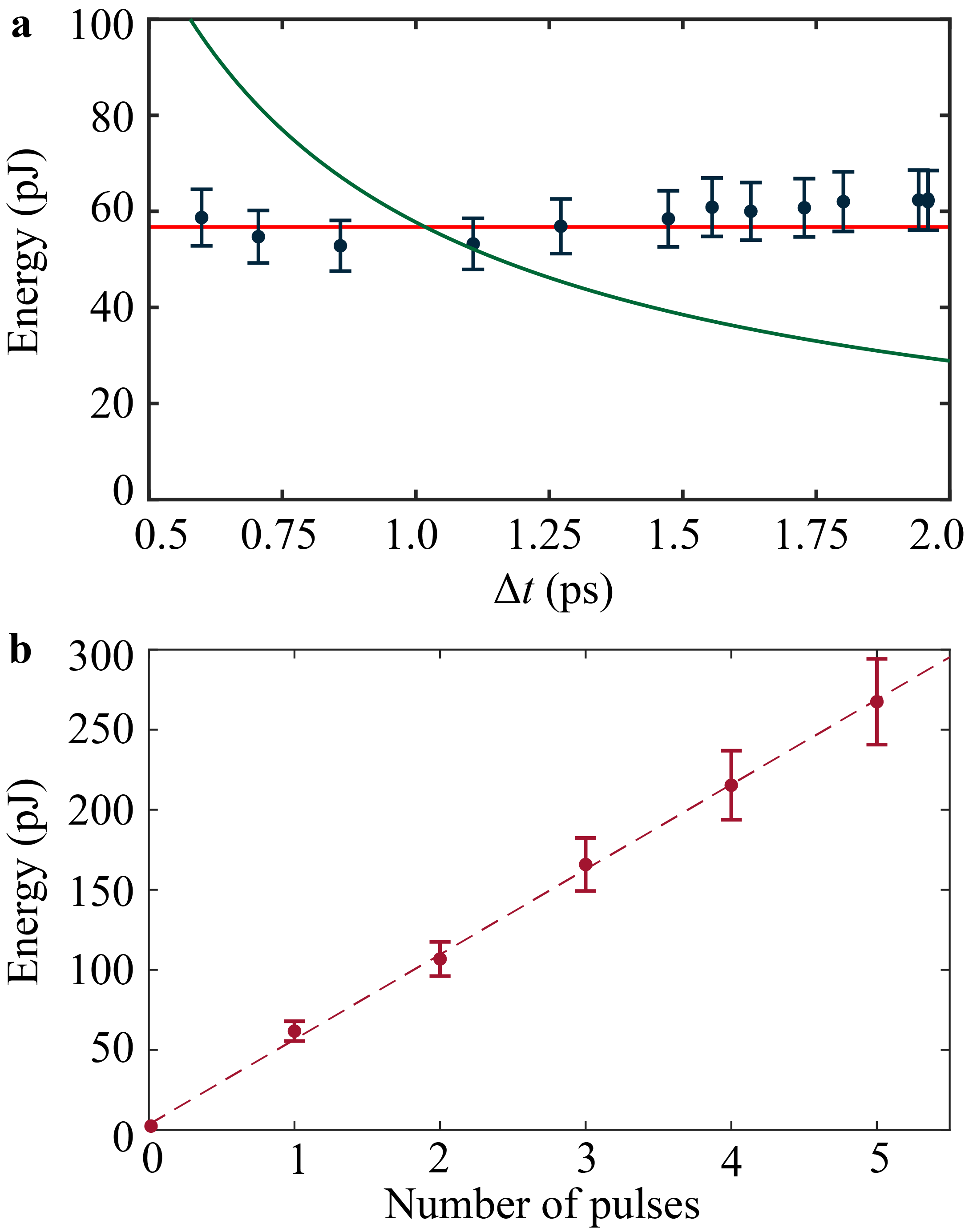}%
\caption{\textbf{Measured energy-width scaling properties of Hilbert-NLS solitons}. (a) Measured pulse energy $E$ versus pulse duration $\Delta t$ (blue circles) for $|\beta_1| = 30~\rm{ps~km^{-1}}$. The red line shows the predicted energy from Eq.~\eqref{eq:scaling_K}. The green line shows the energy scaling of the conventional NLS soliton for our cavity fibre parameters (see Methods). (b) Measured energy versus number of Hilbert-NLS solitons co-propagating in the cavity.}
\label{fig:energy}
\end{figure}

\subsection{Soliton energy-dependence on $\beta_1$ at $\alpha=1$} \label{sec:varying_beta_1}

Equation~\eqref{eq:scaling_K} also predicts that the pulse energy is linearly proportional to the magnitude of the dispersion coefficient $\beta_1$, which can be arbitrarily adjusted by the intracavity pulse-shaper \cite{Runge_2020}. We repeat the measurements from Section~\ref{sec:fixed_beta_1} but for different values of the dispersion coefficient $\beta_1$. We measured the energy for $n_p = 2$ and $n_p = 5$ solitons and divided this energy by $2$ and $5$, respectively. The results of this are shown in Fig.~\ref{fig:beta_1_energy} and are in excellent agreement with the prediction from Eq.~\eqref{eq:scaling_K}. In particular, the line drawn through the data intersects the vertical axis close to the origin, as required. All of the results presented in this section confirm that the solitons observed in this work correspond to the case $\alpha = 1$ and are governed by the Hilbert-NLS equation (Eq.~\eqref{eq:HNLSE}). 

\begin{figure}[ht!]
\centering
\includegraphics[width=7.5cm]{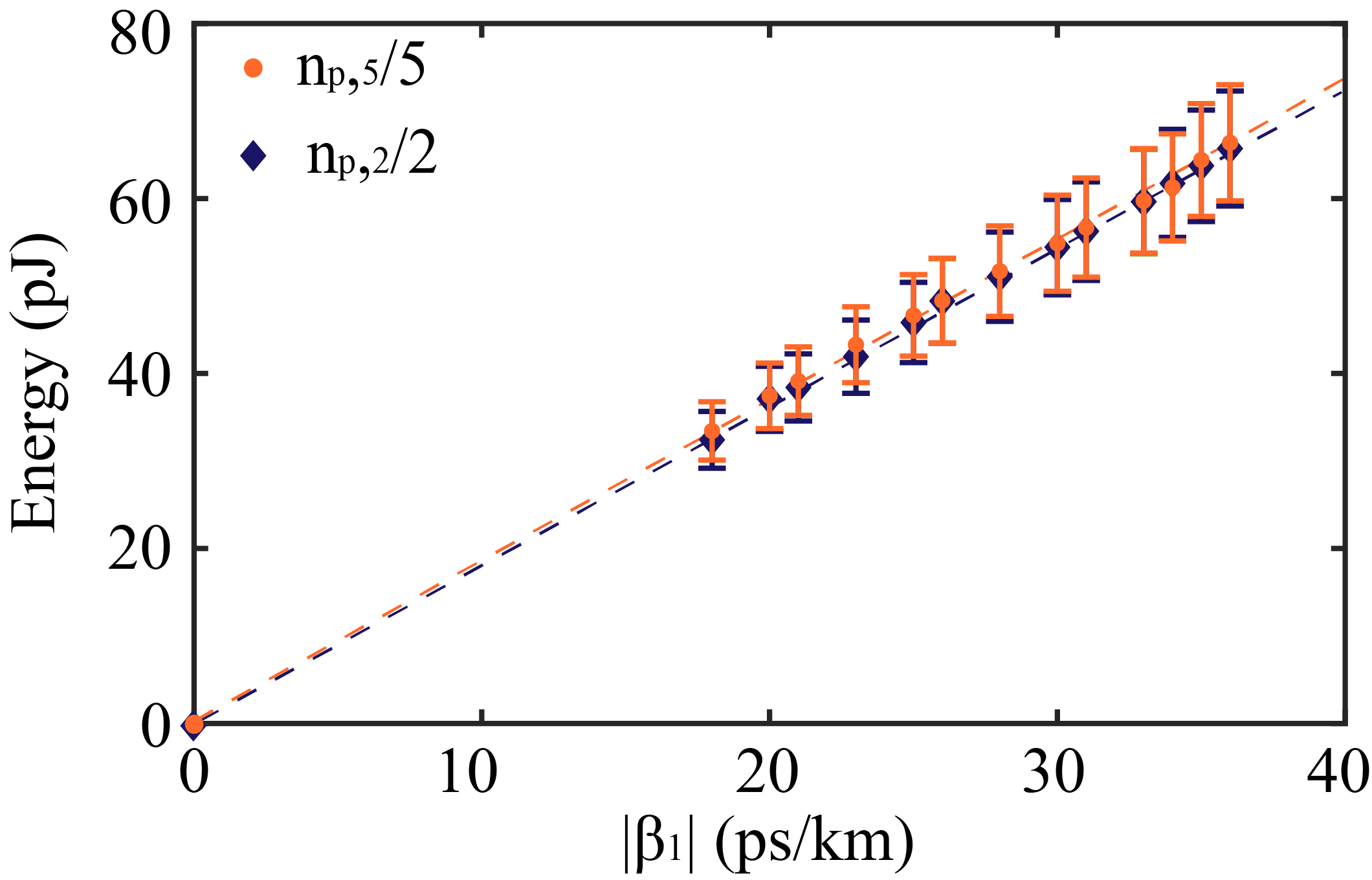}%
\caption{\textbf{Hilbert-NLS soliton energy dependence on $\beta_1$}. Average energy per soliton with  $n_p = 5$ (orange circles) and $n_p = 2$ (blue diamonds) Hilbert-NLS solitons in the cavity versus dispersion coefficient $\beta_1$. The straight lines are linear fits of the data.}
\label{fig:beta_1_energy}
\end{figure}

\subsection{Solitons for $\alpha\neq1$}\label{sec:other_alpha}

Finally, we report the generation of other fractional NLS solitons for L\'evy indices $\alpha\neq1$. We use the same experimental approach as discussed above and apply a phase profile $\phi(\omega) = -|\beta_{\alpha}||(\omega-\omega_0)|^{\alpha}L$, where $|\beta_{\alpha}|$ are constant coefficients. We consider the specific cases $\alpha = \sfrac{5}{4}$ and $\sfrac{\pi}{2}$. We again apply a small negative quartic dispersion to improve the soliton stability, even though the Vakhitov-Kolokolov criterion \cite{Vakhitov_1973} does not indicate unstable behaviour. 

The measured spectra  (solid blue line) corresponding to $\alpha = \sfrac{5}{4}$ and $\sfrac{\pi}{2}$ are shown in Figs.~\ref{fig:alpha}(a) and (b) respectively, and are again in very good agreement with the numerically predicted profiles (red dashed) and previous numerical studies~\cite{Malomed_2024, Wang_2024}. When no phase is applied by the pulse shaper, the laser operates in the conventional NLS soliton regime ($\alpha = 2$). The intrinsic anomalous negative second-order dispersion of the fibre balances the Kerr nonlinearity (see Methods) and the output pulses exhibit the well-known hyperbolic secant spectrum, as seen for comparison in Fig.~\ref{fig:alpha}(c). These measured spectra show that the discontinuous derivative in Fig.~\ref{fig:spectrum}(a) ``softens'' as $\alpha\rightarrow2$ for which case the spectrum is smooth. The associated calculated temporal intensity profiles of these solitons are shown in Supplementary Information. 

\begin{figure}[ht!]
\centering
\includegraphics[width=8cm]{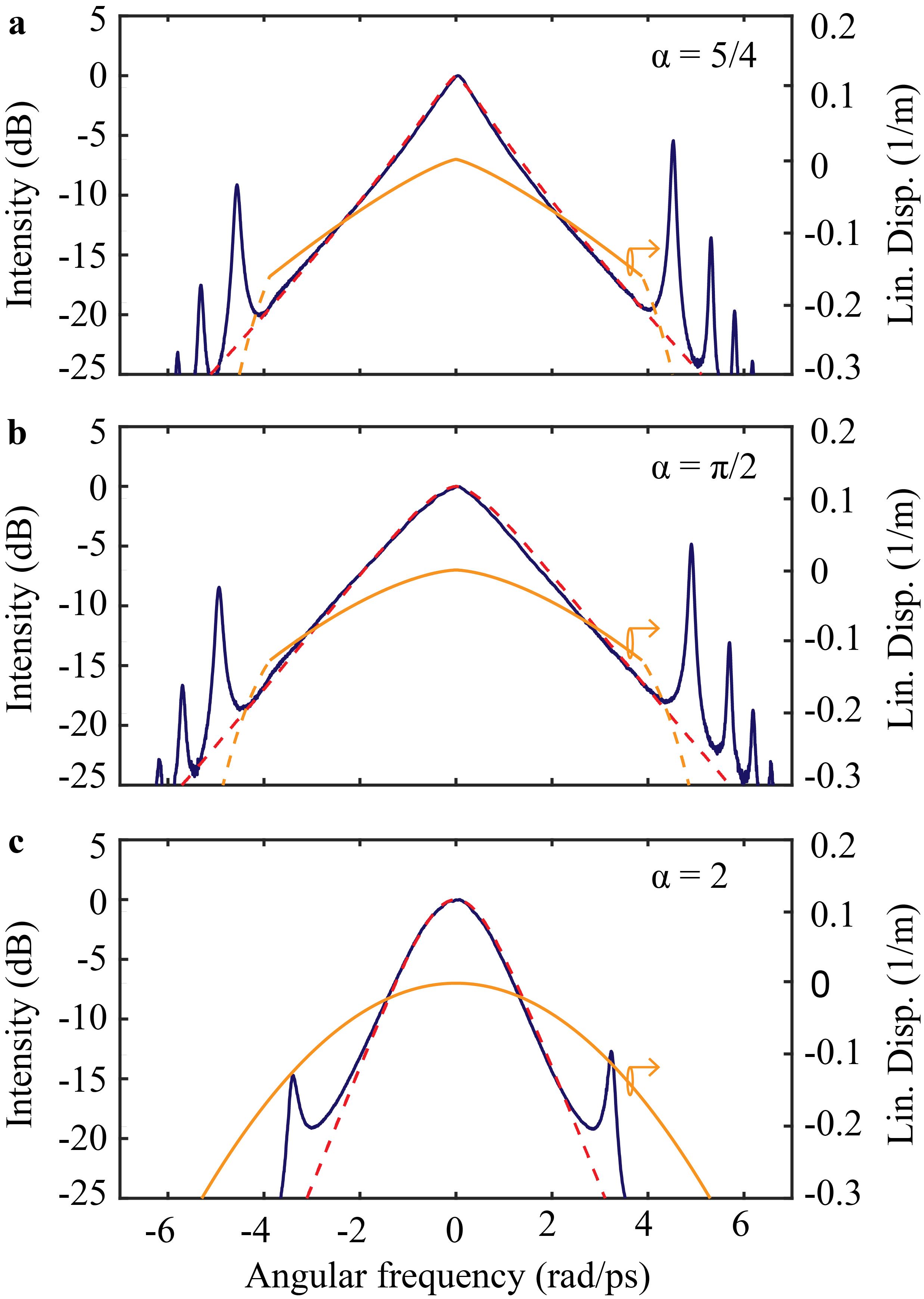}%
\caption{\textbf{Fractional NLS solitons for different values of L\'evy indices $\alpha$.} Measured (solid blue line) and numerically calculated (red dashed) output spectra for (a) $\alpha = \sfrac{5}{4}$, (b) $\alpha = \sfrac{\pi}{2}$, and (c) $\alpha = 2$. The dispersion coefficients for (a) and (b) are set as $\beta_{\alpha} = -30~\rm{ps^{\alpha}~km^{-1}}$ and $-15~\rm{ps^{\alpha}~km^{-1}}$, respectively. The solid curve corresponds to  $\beta=-|\beta_{\alpha}||(\omega-\omega_0)|^{\alpha}$, whereas the dashed curve corresponds to the quartic dispersion included to provide stability. For (c) $\beta_2 = -21.4~\rm{ps^2~km^{-1}}$ corresponds to the second-order dispersion coefficient of single-mode fibre (see Methods). The orange curves show the net-cavity dispersion.}
\label{fig:alpha}
\end{figure}

These results confirm the versatility of our experimental setup and its ability to generate the solutions to a large range of nonlocal and nonlinear wave equations. 
 
\section{Discussion and conclusions}\label{sec:outro}

We have provided the first experimental realization of fractional solitons by harnessing the interaction between fractional dispersion, associated with a fractional Laplacian, and the Kerr nonlinearity in a fibre laser cavity.  
We focused on solitons with the dispersion relation Eq.~\eqref{eq:disp_rel}, and for which the solitons are described by the Hilbert-NLS equation~\eqref{eq:HNLSE}. In a sense this case, for which $\alpha=1$ and the conventional NLS, for which $\alpha=2$ are extremes in the usual interval $0\leq\alpha\leq2$ since for $\alpha<1$ the solitons are unstable. However, we show 
in Section~\ref{sec:other_alpha} 
that our experimental approach is flexible enough to dial in any value $1\leq\alpha\leq2$ and that the resulting solitons share features of those of the extreme cases.

Fractional Laplacians are nonlocal operators, and indeed this property enters Eq.~\eqref{eq:HNLSE} through the presence of the Hilbert transform. This leads to the discontinuity in the derivative of the spectrum, which in turn leads to the non-exponential asymptotic temporal behavior of the solutions. The non-locality makes it difficult to find analytic solutions of Eq.~\eqref{eq:stationary} since approaches that rely on local or asymptotic expansions cannot be used. Even though closed-form solutions to the Benjamin-Ono equation  \cite{Benjamin_1967},
which also involves a Hilbert transform and which is somewhat related to Eq.~\eqref{eq:stationary}  \cite{Roudenko_2021}, are known, these solutions do not carry over to the present case.

Experiments on marginally stable solitons have been carried out before \cite{Parades_2004}. In that previous work, the solitons were stabilized by a lattice potential. Our approach here is different and relies on subtle changes to dispersion relation Eq.~\eqref{eq:disp_rel}. Even though this leads to slight changes in the generated solitons, the overall effects---while most clear in our pulse energy measurements (see Fig.~\ref{fig:energy})---are nonetheless quite modest.

In conclusion, we report the generation and full characterization of nonlocal solitons that satisfy the Hilbert-NLS equation, a special case of the fractional NLS equation. These solitons have striking properties which we now summarise. 
\begin{itemize}[wide=0pt,noitemsep,topsep=0pt] 
\item The soliton spectrum has a discontinuous derivative, which reflects the discontinuous derivative of the dispersion relation Eq.~\eqref{eq:disp_rel}. In turn, this discontinuous derivative leads to non-exponential asymptotic decay in time. This reflects the nonlocal properties of the Hilbert transform that  enters the fractional Laplacian. 
\item These solitons have a very small time-bandwidth product, much smaller than that of conventional solitons. This can be attributed to the compact spectrum, which is a consequence of the discontinuous derivative. Even though the algebraic decay implies a broad spectrum, this is not yet apparent at the relatively high intensities that determine the solitons' FWHM. 
\item The soliton energy is independent of their peak power or width. This property is equivalent to their marginal stability according to the Vakhitov-Kolokolov criterion \cite{Vakhitov_1973}. In other words, this marginal stability implies that the soliton energy is independent of its width.
\end{itemize} 

Even though our experiments were carried out in an optics context, our results are universal and agnostic to the particular physical embodiment of the governing equation. As such, our results can act as a starting point for a new wave of experimental investigations of linear and nonlinear wave phenomena in media with complicated dispersion relations.

\section*{Methods}

\textbf{Hilbert-nonlinear Schr\"odinger equation}. 
The nonlinear Schr\"odinger (NLS) equation with generalized dispersion and with a cubic nonlinearity takes the general form 
\begin{equation}
    i\frac{\partial\varphi}{\partial z}+{\cal O}_D(\varphi)+\gamma |\varphi|^2\varphi=0,
\label{eq:proto_HNLSE}
\end{equation}
where ${\cal O}_D$ is an operator describing the dispersion. For the conventional NLS equation the dispersion relation takes the form $\beta=\sfrac{1}{2}~\beta_2\omega^2$ \cite{Kivshar}, where $\beta$ is the wavenumber, $\omega$ is the frequency and $\beta_2$ is a coefficient that expresses the strength of the effect. In this case ${\cal O}_D=-(\sfrac{\beta_2}{2})(\sfrac{\partial^2}{\partial t^2})$. Substituting $\varphi\sim e^{i(\beta z-\omega t)}$ and ignoring the nonlinear term then gives the correct dispersion relation. 

For the derivation of Eq.~(\ref{eq:HNLSE}) in the main text we require the Hilbert transform \cite{King}
\begin{equation}
    \HH(u(t))=\frac{1}{\pi}~{\cal P}\!\int_{-\infty}^\infty \frac{u(\tau)}{t-\tau}d\tau,
\label{eq:Hilbert_convolve}
\end{equation}
where ${\cal P}$ denotes the principal value. The Hilbert transform has the property that \cite{King}
\begin{equation}
    \HH(u(t))= i{\cal F}^{-1}({\rm sgn}(\omega)U(\omega)),
\label{eq:Hilbert}
\end{equation}
where ${\cal F}^{-1}$ is the inverse Fourier transform. Thus the effect of the Hilbert transform is to reverse the signs of the amplitudes of the spectral components with negative frequencies (the factor $i$ guarantees that the Hilbert transform of a real function is real). Since $|\omega|=\omega~{\rm sgn}(\omega)$ where ${\rm sgn}$ is the sign function it is not surprising that Eq.~(\ref{eq:HNLSE}) involves a Hilbert transform. 

We can check that Eq.~(\ref{eq:HNLSE}) is consistent with the dispersion relation in Eq.~(\ref{eq:disp_rel}) by again using substituting $\varphi\sim e^{i(\beta z-\omega t)}$ and ignoring the nonlinear term. With this ansatz in Eq.~(\ref{eq:HNLSE}) and using $\HH(e^{-i\omega t})=i~{\rm sgn}(\omega) e^{-i\omega t}$ \cite{King}, we immediately find Eq.~(\ref{eq:disp_rel}).

\textbf{Soliton scaling and instability}.
Equation~(\ref{eq:stationary}) in the main text has a number of scaling properties according to which the solutions must satisfy 
\begin{equation}
    u(t)=\sqrt{\frac{\mu}{\gamma}}~ f\left[\left(\frac{\mu}{|\beta_1|}\right)t\right],
\label{eq:scaling}
\end{equation}
where $f$ is a function that needs to be found numerically. To illustrate this, consider the following example: suppose we have a solution of  Eq.~(\ref{eq:stationary}). Then the $\mu$ dependence in Eq.~\eqref{eq:scaling} ensures that if the $u$ is shortened by a factor $\eta$ and the amplitude increases by a factor $\sqrt\eta$, 
then every term in Eq.~(\ref{eq:stationary}) increases by a factor $\eta^{3/2}$ in magnitude, so that the resulting function is also a solution. 

Now the pulse energy is given by $\int u^2(t) dt$. Carrying out this integration we then immediately find Eq.~(\ref{eq:scaling_K}). The fact that $\mu$ cancels out indicates that, for fixed $\beta_1$ and $\gamma$, the pulse energy does not depend on the pulse width or peak power.

The condition for stability of a soliton has been discussed by many authors. Boulenger \textit{et al.} \cite{Boulenger_2016}, for example,  express it in terms of a scaling index $s_c$
\begin{equation}
    s_c=\frac{N}{2}-\frac{\alpha}{\sigma},
\end{equation}
where $N$ is the dimensionality, $\alpha$ is defined in the main text and $\sigma$ describes the nonlinear term through $|\varphi|^{2\sigma}\varphi$. For $s_c<0$ the system is subcritical, for $s_c>0$ it is supercritical, whereas for $s_c=0$ it is critical. In our case $N=1$, $\alpha=\sfrac{1}{2}$ and $\sigma=1$ and so $s_c=0$, corresponding to the critical case. By comparison, for the conventional NLS equation $N=1$, $\alpha=1$ and $\sigma=1$, so $s_c=-\sfrac{1}{2}$, corresponding to the stable, subcritical case. The finding that $s_c=0$ follows from the scaling relation derives above and is equivalent to the observation that that the energy $E$ does not depend on $\mu$. 


\textbf{Experimental laser cavity.}
The total laser cavity of length $L = 18.2~\rm{m}$ is made of sections of standard telecommunication single-mode fibre (SMF-28) and a short section of erbium-doped fibre ($1~\rm{m}$). SMF-28 has a nonlinear coefficient $\gamma = 1.3~\rm{W^{-1}~m^{-1}}$ at $1560~\rm{nm}$. The dispersion coefficients are $\beta_2 = -21.4~\rm{ps^2~km^{-1}}$ and $\beta_3 = 0.12~\rm{ps^3~km^{-1}}$. As the length of erbium-doped fibre is much shorter than the total cavity length and the mode-field diameter and numerical aperture are close to that of SMF-28, we assumed similar dispersion coefficients and nonlinear coefficients.

The intracavity pulse shaper is programmed to induce a total spectral phase that: (i) compensates the second- and third-order dispersion introduced by the optical fibres, and (ii) generates a fractional Laplacian following Eq.~\eqref{eq:disp_rel}. To stabilize the solitons, quartic dispersion is applied at the spectral pulse edges, beginning from the intersection points of the two dispersion curves (i.e., $-|\beta_4|(\omega-\omega_0)^4/24 = -|\beta_{\alpha}||(\omega-\omega_0)|^{\alpha}$) extending towards the pulse edges. The values of $\beta_4$ are chosen to ensure long-term stability and a high intensity contrast (e.g., higher than 20 dB) between the central and the intersection points. Thus, $\beta_4 = -40$ ps$^4$~km$^{-1}$ is used in Fig.~\ref{fig:spectrum} and $\beta_4 = -20$ ps$^4$~km$^{-1}$ for the other figures.

\textbf{Phase-resolved spectro-temporal characterization.} The output pulses were sent into the FREG setup. The pulses were split into two branches by a 70/30 fibre-coupler; 30\% of the output power was sent to a branch with a variable delay before being detected by a fast photodiode and transferred to the electrical domain. This electrical signal drove an intensity modulator that gated the optical pulses from the 70\% branch of the fibre-coupler. Using an optical spectrum analyser, we measured the spectra as a function of the delay to generate the associated optical spectrograms. The spectra were recorded over a $ \rm{nm}$ spectral bandwidth with a $\Delta\lambda = 0.1~\rm{nm}$ resolution, and the temporal delay was scanned over $40~\rm{ps}$ with a $0.2~\rm{ps}$ temporal step. The experimental spectrograms were then deconvolved with a blind deconvolution numerical algorithm ($1024\times1024$ grid; retrieval errors $\leq0.001$) to retrieve the pulse amplitude and phase in the temporal domain.

\section*{Supplementary Information}

\subsection*{Experimental setup}

The laser setup is shown schematically in Fig.~\ref{fig:laser_setup} and is similar to that of Refs.~\cite{Runge_2020, Lourdesamy_2022} for studying the effects of higher-order dispersion. It operates at $1560~{\rm nm}$ with gain supplied by erbium-doped fibre which is pumped by two laser diodes operating at $980~\rm{nm}$, coupled into the cavity by two wavelength division multiplexers. The cavity includes an optical isolator to ensure unidirectional propagation of light. Two fibre polarization controllers combined with a fibre polariser effectively act as a passive saturable absorber for mode-locking to achieve short pulses \cite{Fermann_1993}. The remainder of the cavity consists of standard SMF-28 single-mode fibre. The output coupler extracts 50\% of the cavity light. A programmable spectral pulse shaper (WaveShaper) is included inside the cavity, which allows a frequency-dependent phase $\varphi(\omega)$ to be applied~\cite{Baxter_2006, Roelens_2008}. We consider
\begin{widetext}
\begin{equation}
    \varphi(\omega) =L\left[ -\left\{ \frac{\beta_{2,{\rm SMF}}}{2!}(\omega-\omega_0)^2 + \frac{\beta_{3,{\rm SMF}}}{3!}(\omega-\omega_0)^3 \right\}+ |\beta_{\alpha}||(\omega-\omega_0)|^{\alpha} \right], \label{eq:laser_dispersion}
\end{equation}
\end{widetext}

where $\omega$, $\omega_0$ are as defined in the main text. Additionally, $L = 18.2$~m is the approximate cavity length ~\cite{Lourdesamy_2022}, while $\beta_{2,{\rm smf}} = -21.4~{\rm ps}^2~{\rm km}^{-1}$ and $\beta_{3,{\rm smf}}=0.12~{\rm ps}^3~{\rm km}^{-1}$ are the leading dispersion coefficients for SMF-28 fibre ~\cite{Runge_2020,Hammani_2011}. As per Eq.~\eqref{eq:laser_dispersion}, these are canceled out using the WaveShaper. The final term in Eq.~\eqref{eq:laser_dispersion} represents the applied net average dispersion across the cavity, as illustrated in Fig.~\ref{fig:laser_setup}.

\begin{figure*}[ht!]
\centering
\includegraphics[width=13cm]{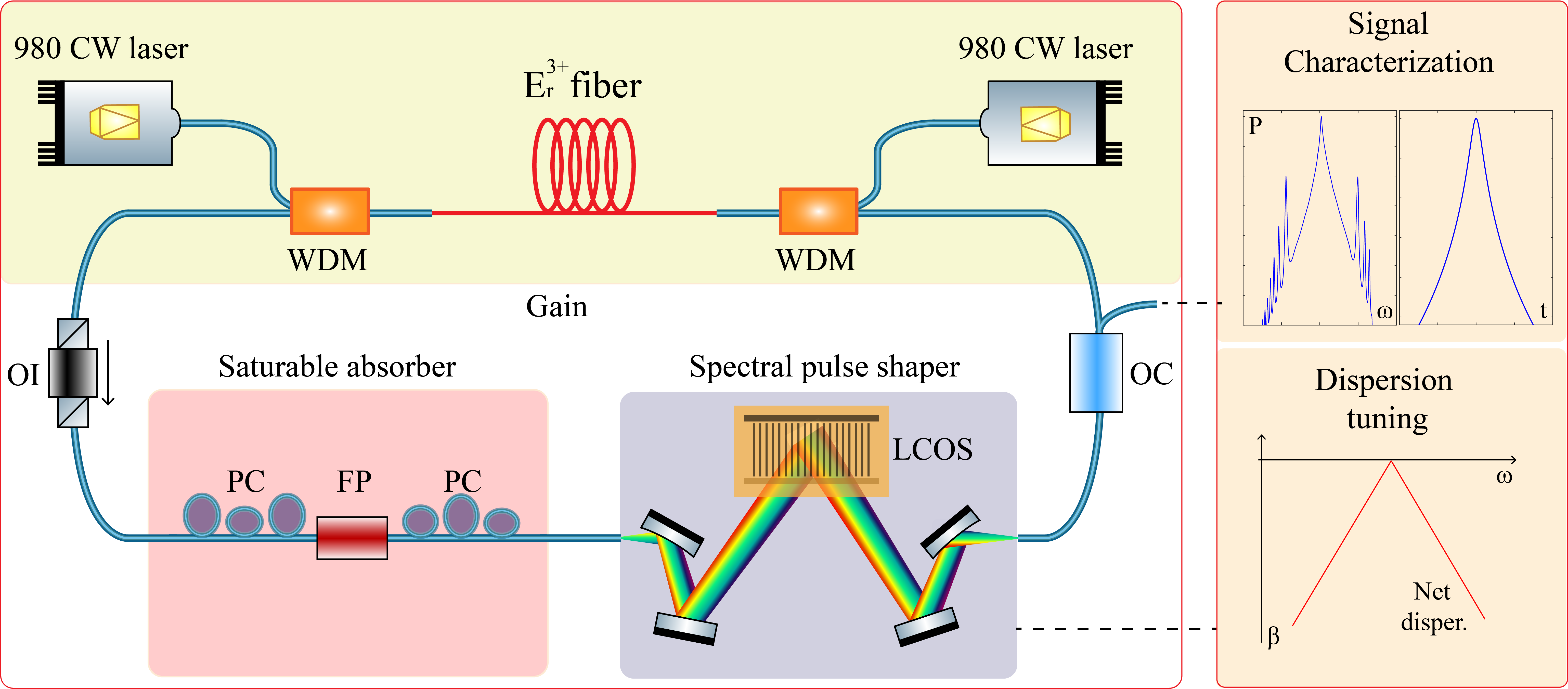}
\vskip -1mm
\caption{Schematic of the experimental setup to generate fractional NLS solitons. $E^{3+}_r$ fiber, erbium-doped fibre; WDM, wavelength division multiplexers; OI, optical isolator; PC, polarization controller; FP, in-line fibre polarizer; LCOS, liquid crystal on silicon; OC, output coupler.}
\label{fig:laser_setup}
\end{figure*}

\subsection*{Laser numerical simulation model}

Laser numerical simulations are based on the NLS Eq. (5) in the main text with a gain term: 
\begin{equation}
   \frac{\partial A}{\partial z}-i{\cal O}_D(A) 
=\frac{g}{2}A+i\gamma A|A|^2.
    \label{GNLSE}
\end{equation}

with the dispersion operator defined as:
\begin{equation}
{\cal O}_D = \sum_{k}\frac{\beta_k}{k!}\left(i\frac{\partial}{\partial \tau}\right)^k,
    \label{Dop}
\end{equation}
for $k = 2,3$. $\beta_k$ is the $k^{th}$ order of dispersion. The gain in the doped fibre section is calculated using~\cite{Oktem_2010}:
\begin{equation}
   g = \frac{g_0}{1+E(z)/E_{\rm sat}}.
    \label{gain}
\end{equation}
where $g_0 = 3.45$ is the small-signal gain (corresponding to 30 dB in power and non-zero only in the doped-fibre section), $E(z) = \int|A(z,T)|^2$ is the pulse energy and $E_{\rm sat}$ is the saturation energy, which is adjusted to simulate changing the pump power. We multiply $g(z)$ with a Lorentzian profile of $50$ nm width to form the finite gain bandwidth $g(z,\omega_0)$. The saturable absorber is modelled by a transfer function that describes its transmittance
\begin{equation}
   T(\tau) = 1-\frac{q_0}{1+P(\tau)/P_0}.
    \label{SA}
\end{equation}
where $q_0$ is the unsaturated loss of the saturable absorber, $P(\tau) = |A(z,\tau)|^2$ is the instantaneous pulse power and $P_0$ is the saturation power. The spectral pulse-shaper is modelled by multiplying the electric field by a phase following the expression in Eq.~\ref{eq:laser_dispersion} in the spectral domain. The insertion losses ($\approx 5.6$ dB) of the spectral pulse-shaper are also taken into account in the simulations. Our numerical model is solved with a standard symmetric split-step Fourier method algorithm \cite{Agrawal_NFO}. The dispersion and gain contributions are calculated in the frequency domain, whilst the nonlinear term is calculated in the time domain. For our simulations we have used an initial field composed of Gaussian random noise multiplied by a sech shape in the time domain. The same stable solutions are reached for different initial noise fields.

\subsection*{Effects of quartic dispersion on the spectral and temporal profiles of Hilbert-NLS solitons} 

To experimentally address the marginal stability of Hilbert-NLS solitons we add a small negative quartic dispersion at frequencies far from the central frequency $\omega_0$. The dispersion profile used in the experiments (solid orange) and the dispersion profile following Eq.~(1) in the main text (dashed green) are shown in Fig.~\ref{fig:JY_calcu}(a). The numerically calculated Hilbert-NLS soliton spectral intensities associated to these two dispersion profiles are shown in Fig.~\ref{fig:JY_calcu}(a) and only slightly differ at low intensity away from the central frequency. The corresponding temporal profiles are shown in Fig.~\ref{fig:JY_calcu}(b). When the negative quartic dispersion is included, the temporal profile exhibits small oscillations arising from the discontinuous spectrum, and is in good agreement with the measured temporal profile shown in Fig.~2(b) in the main text.

\begin{figure*}[h!]
\centering
\includegraphics[width=13cm]{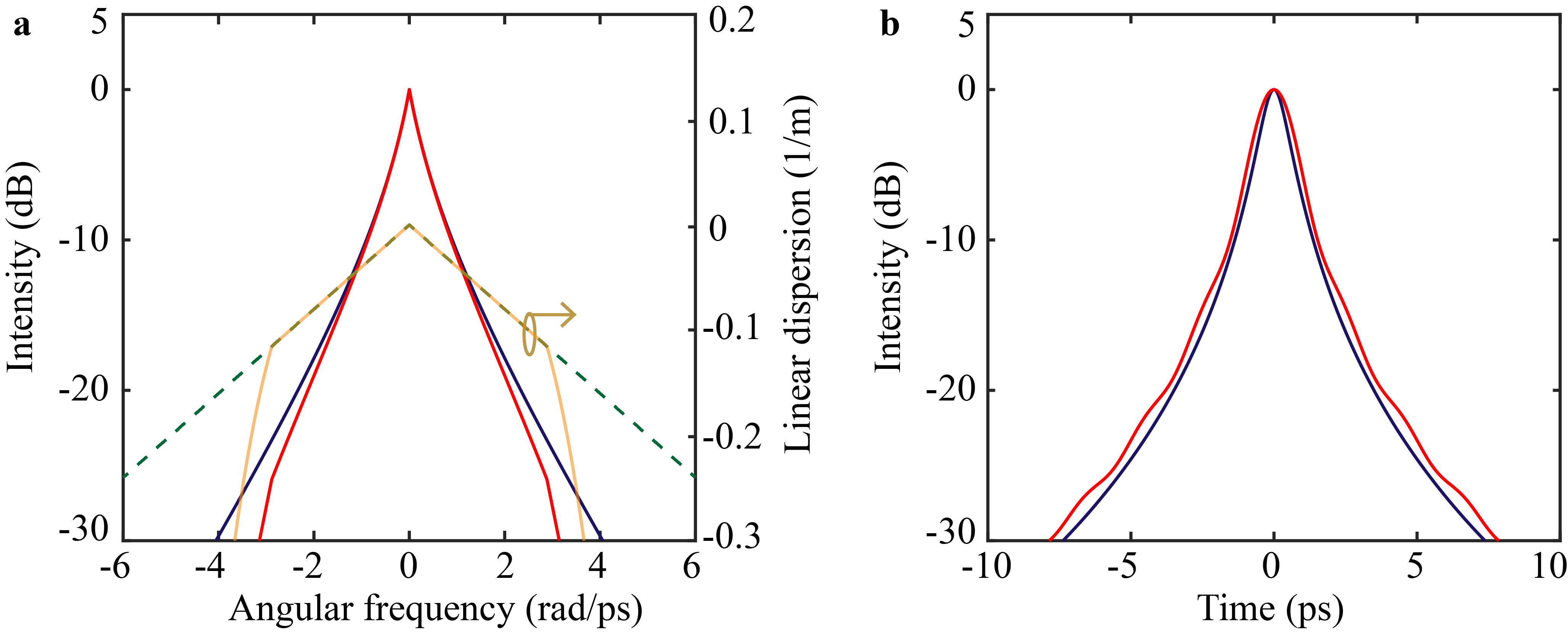}
\vskip -1mm
\caption{(a) Numerically calculated spectral profiles of Hilbert-NLS solitons with (red) and without (blue) quartic dispersion. The green dash and orange line, respectively, show linear dispersion $\beta(\omega)=-|\beta_1||(\omega-\omega_0)|$ and the dispersion with quartic $\beta(\omega)=\beta_4(\omega-\omega_0)^4/24$ applied to the spectral pulse edges, in which $|\beta_1|=40~{\rm ps}~{\rm km}^{-1}$ $\beta_4=-40~{\rm ps}^4~{\rm km}^{-1}$. (b) Corresponding temporal intensity profiles.}
\label{fig:JY_calcu}
\end{figure*}

\subsection*{Multi Hilbert-NLS soliton regime characterization} 

For large pump power, multiple Hilbert-NLS soliton can co-propagate in the laser cavity. We determine the number of solitons $n_p$ existing in the cavity by recording the laser output pulse train using a $1~\rm{GHz}$ photodiode and a $1~\rm{GHz}$ real-time oscilloscope. Examples of recorded pulse trains for $n_p = 1$, $3$ and $5$ solitons are shown in Fig.~\ref{fig:OSCI}. The separation between the different solitons is in the order of tens of nanoseconds, longer than the photodetector response and orders of magnitude longer than the pulse duration ($\approx 2~\rm{ps}$). We numerically calculate the energy of each soliton by integrating the recorded time trace and found results similar to Fig.~2(a) in the main text, i.e., the energy of each soliton is equal and consistent with Eq.~(4) in the main text.

\begin{figure*}[h!]
\centering
\includegraphics[width=13cm]{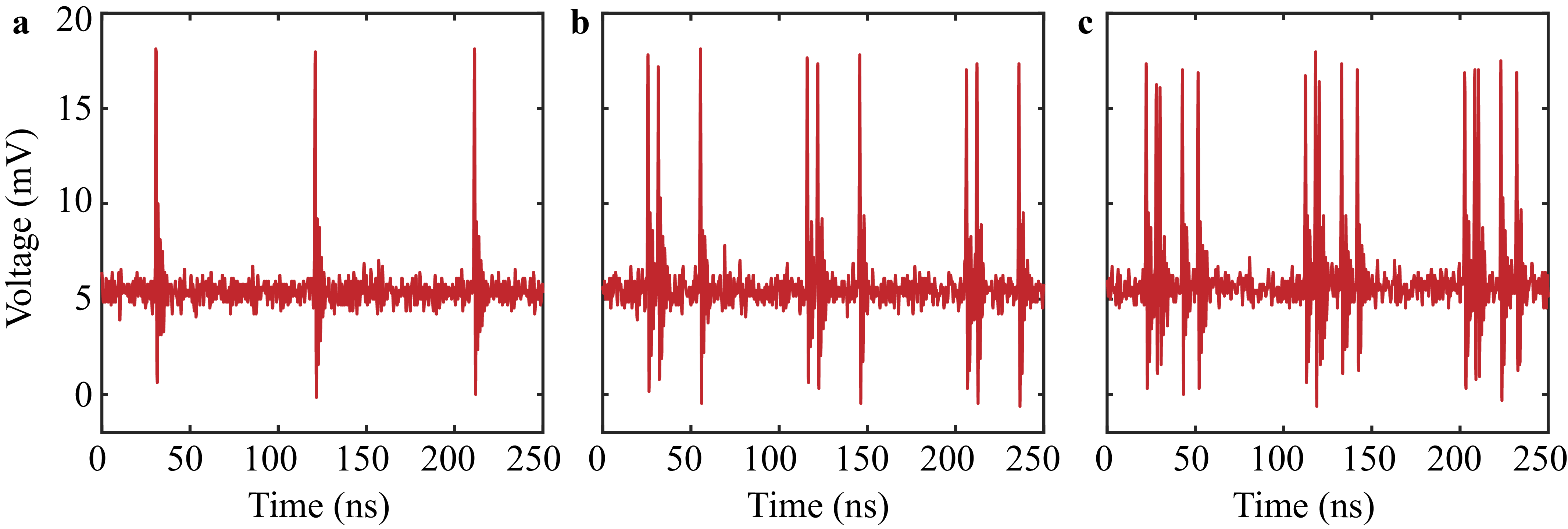}
\vskip -1mm
\caption{Recorded output pulse trains for (a) $n_p=1$, (b) $n_p=3$, and (c) $n_p=5$.}
\label{fig:OSCI}
\end{figure*}

These results is also confirmed by the average output spectra for $n_p = 1$, $3$ and $5$ solitons are shown in Fig.~\ref{fig:1-5solitons}. For all cases, the output spectral bandwidths remain the same as well as the spectral position of the Kelly side-bands, as shown in Fig.~\ref{fig:1-5solitons}(a), suggesting that the nonlinear phase shift $\mu$ and therefore pulse energy remains constant \cite{Kelly_1992}. Finally, we measured the total energy of the solitons for different pump powers and different number of solitons $n_p$. The results in Fig.~\ref{fig:1-5solitons}(b) show that (i) the total energy remains approximately constant while the spectral bandwidth (pulse duration) increases (decreases) with increasing pump power, consistent with Fig.~2(a) in the main text; and (ii) the total energy takes discrete values that are multiple of the single Hilbert-NLS soliton energy given by Eq.~(4) in the main text ($\approx 52~\rm{pJ}$).

\begin{figure*}[h!]
\centering
\includegraphics[width=13cm]{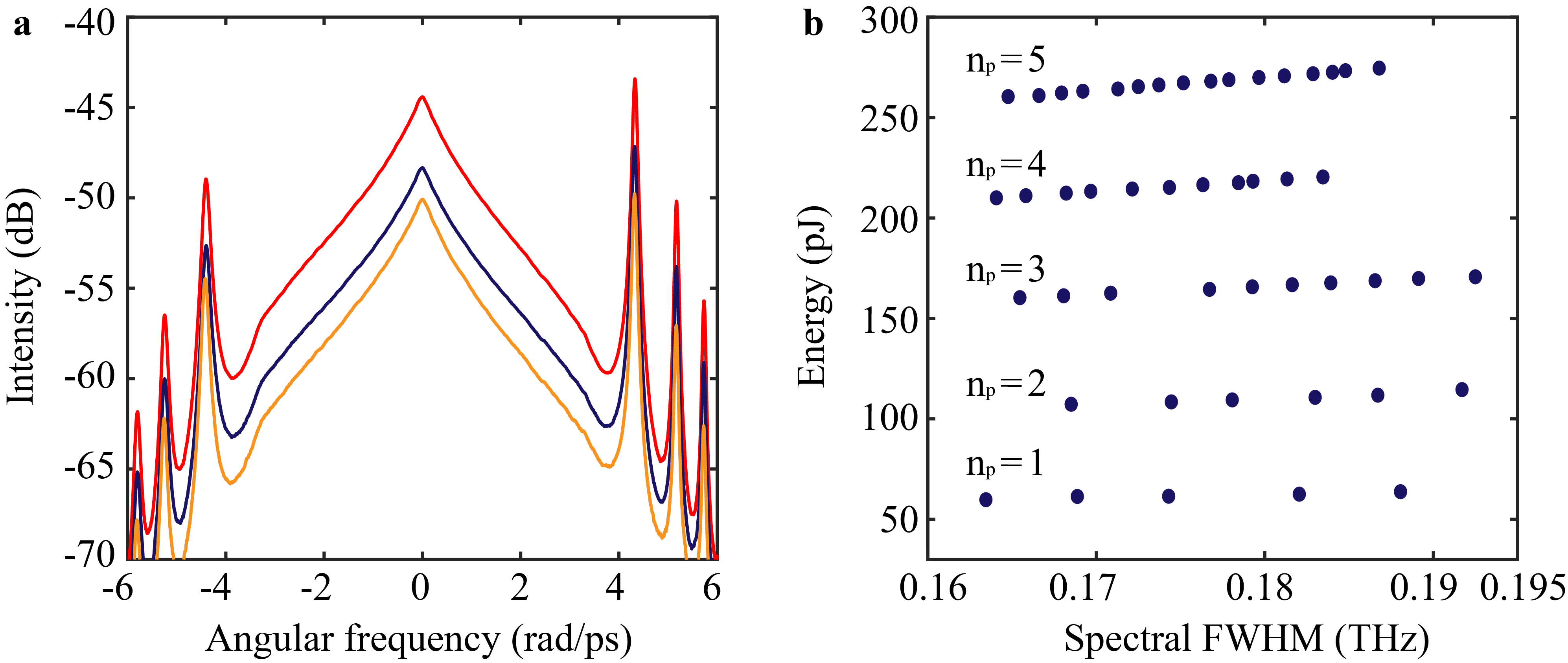}
\vskip -1mm
\caption{Measured output average spectra and energy for different numbers of solitons in the cavity with $|\beta_1|=30~{\rm ps}~{\rm km}^{-1}$. Spectra corresponding to 5 (red line), 2 (blue line), and 1 (orange line) solitons. These spectra have a similar value of spectral FHWM. (b) Energy with soliton number from 1 to 5 over spectral FHWM.}
\label{fig:1-5solitons}
\end{figure*}

\subsection*{Temporal profiles of solitons at $\alpha\neq1$}

\begin{figure*}[ht!]
\centering
\includegraphics[width=13cm]{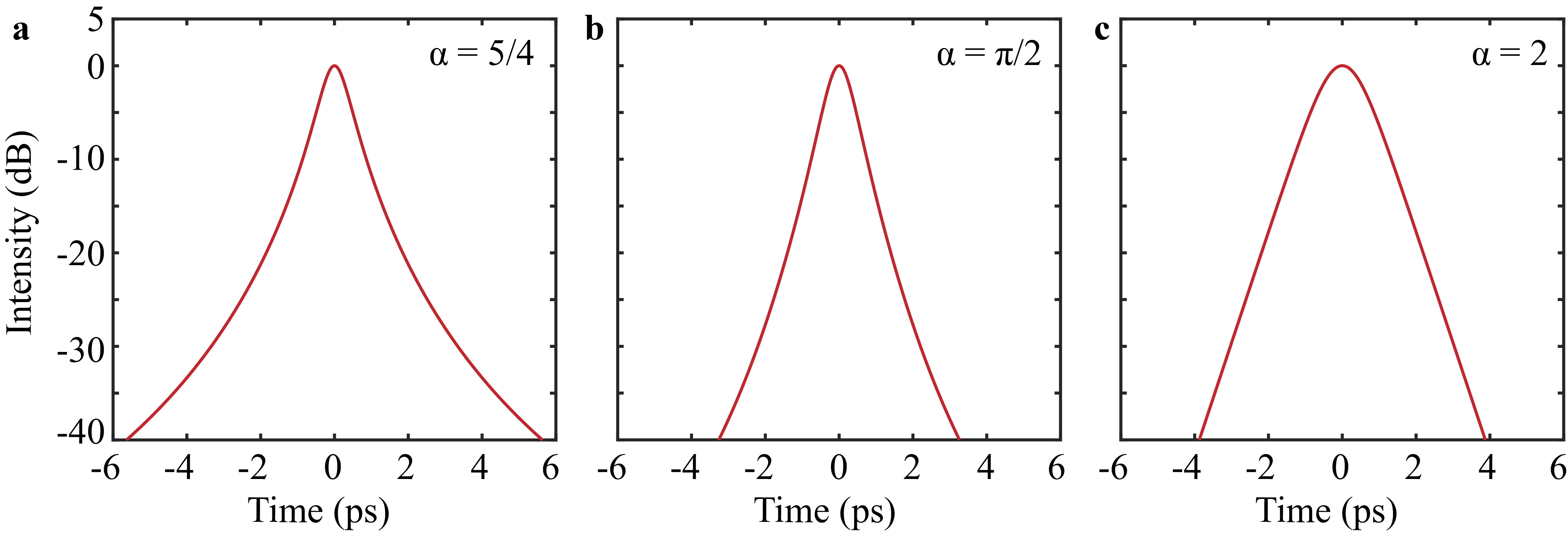}
\vskip -1mm
\caption{Numerically calculated temporal profile for (a) $\alpha = \sfrac{5}{4}$, (b) $\alpha = \sfrac{\pi}{2}$, and (c) $\alpha = 2$.}
\label{fig:general_tem}
\end{figure*}
The numerically calculated intensity temporal profiles corresponding to the fractional soliton spectra shown in Fig.~5 in the main text, are shown in Fig.~\ref{fig:general_tem}. For $\alpha = 5/4$ and $\pi/2$, the calculated corresponding time bandwidth product $\Delta t\Delta f$ are $0.13$ and $0.22$, respectively. Similar to the case for $\alpha = 1$ discussed in the main text, the temporal profile of these fractional solitons decays sublinearly on a logarithmic scale, indicating a slower-than-exponential decay. Finally for $\alpha = 2$, corresponding to the conventional NLS soliton, the temporal intensity profile follows an hyperbolic secant shape \cite{Kivshar}.

\section*{Data availability}
The data that support the plots in this paper and other findings of this study are available from the corresponding author upon reasonable request.

\section*{Acknowledgements}
This research was supported by the Australian Research Council (ARC) Center of Excellence in Optical Microcombs for Breakthrough Science (project no. CE230100006), funded by the Australian Government. A.F.J.R. is supported by the ARC Discovery Early Career Researcher Award (DE220100509).  V.T.H., A.F.J.R. \& C.M.deS are also supported by an ARC Discovery Project (DP230102200).

\section*{Author contributions}
V.T.H. performed the experiments and numerical simulations. A.F.J.R. designed the experimental setup. J.W. performed early experiments and numerical simulation. Y.L.Q and M.L performed numerical simulations. A.F.J.R. and C.M.d.S. supervised the overall project. All of the authors contributed to the interpretation of the data and wrote the manuscript.

\section*{Competing interests}
The authors declare no competing interests.


\bibliography{References}

\end{document}